\documentclass{article}

\usepackage{graphicx}
\usepackage{graphpap,geometry}
\usepackage{amsmath}
\usepackage{amsfonts,amssymb}
\usepackage{cite}
\newcommand{\ket}[1]{| #1 \rangle}

\newcommand{\rb}[1]{\left( #1 \right)}

\title{{\bf  Bogoliubov transformations and exact isolated solutions for
simple non-adiabatic Hamiltonians}}
\author{
C. Emary and R. F. Bishop\\
{\it Department of Physics},\\
{\it  University of Manchester Institute of
Science and Technology (UMIST)},\\
{\it P. O. Box 88, Manchester M60 1QD,
United Kingdom }}
\begin{document}
\parindent=0in
\maketitle
\begin{abstract}
\noindent
We present a new method for finding isolated exact solutions of a class
of non-adiabatic Hamiltonians of relevance to quantum optics and allied 
areas.  
Central to our approach is the use of
Bogoliubov transformations of the bosonic fields in the models.  We 
demonstrate the simplicity and efficiency of this method by applying it to
the Rabi Hamiltonian.
\end{abstract}

PACS number(s):  03.65.-w, 42.50.-p, 32.80.-t
\normalsize
\newpage
\section{Introduction}

There exists a class of simple, non-integrable, adiabatic Hamiltonians
of the type that find application as models of light-matter
interactions, for which it is possible to 
find exact isolated solutions. 
Generally these models
involve some atomic system, typically characterised by a simple two-level
(or multi-level) system,  interacting with a number of bosonic fields.
Making the familiar rotating-wave approximation usually renders these models 
completely soluble, but avoiding this approximation maintains the
non-integrability of the models, and gives rise to the possibility of
isolated exact solutions. This
was first demonstrated for the Jahn-Teller model by Judd \cite{ju:dd}, 
and these
solutions are often referred to as Juddian solutions.  Probably the simplest
model for which these solutions have been found is the
Rabi Hamiltonian (RH), which describes a two-level atom interacting with
a single-mode bosonic field via a dipole interaction \cite{al:eb}.  
The Juddian solutions of the RH were first discovered
by  Reik and co-workers \cite{re:nu}, where they were seen to
occur at the level crossings in the energy schema of the system.  This turns 
out to be a general and important feature of these solutions.

Apart from being of interest for what they tell us about the structure 
and symmetries of these models, the Juddian solutions are of considerable 
further value.  Simple quantum optics and related models, 
such as the RH, have long 
been utilised as test cases for various calculational techniques 
\cite{fe:ra, bi:cc, bi:va},
and the possession of exact solutions facilitates their accurate 
assessment.  Furthermore, the existence of isolated exact solutions in 
non-integrable quantum models is also
of interest from the perspective of studying possible quantum chaos 
in such systems \cite{gr:ho, ho:ta}.  In addition, it is hoped that
these exact solutions may serve as useful starting points for 
perturbative treatments of the entire spectra of these models.

In this paper we present a new and more general method for finding 
these isolated exact solutions, which we believe to have several 
advantages over the methods hitherto employed. Judd and Reik, working in the 
Bargmann representation, have used power series and Neumann series
Ans\"{a}tze for the field mode.  Neither of these approaches is particularly
intuitive and the resulting algebra can become complicated.  K\`{u}s and 
Lewenstein \cite{ku:le} have given a more concise approach which, as we 
describe later, is clearly 
related to the method that we describe here.  For models such
as the RH they used Bargmann representation Ans\"{a}tze for 
the field consisting of a finite 
number of bosonic excitations on top of a coherent state.  They have also 
extended their method to some further systems, such as a three-level 
system and  an auto-ionising ion. 

We believe that the method we outline in this paper is both more intuitive and
more efficient that those discussed above, and that it reflects the essential
physics of the systems to a greater degree.  At the heart of the method
is a simple 
canonical transformation of the bosonic field operators of the models.  This 
transformation  suggests the existence of exact solutions in a most direct 
manner.  Our method also has the advantage that it is easy to generalise,
and is readily able to be extended
to ``two-photon'' type interactions, in which two photons are required
to induce an atomic transition \cite{pu:bu}.

The remainder of this paper is organised as follows.  In Sec. \ref{BogT} we 
outline our method for finding the Juddian solutions.  We describe in
some detail the theory of Bogoliubov transformations of a boson mode and 
pay particular attention to their relation
to the coherent and squeezed states.  We then use these transformations 
to investigate the displaced and squeezed harmonic oscillators, to
develop insight into the reasoning behind this approach. In Sec. 
\ref{RHapp} we apply this method to the Rabi Hamiltonian, as an example of 
the use of this method.  We then finish with some conclusions and indications
of further work.

\section{Methodology \label{BogT}}

The models that we consider here consist of an atomic
system interacting with one or more bosonic modes.
Each of these  modes is described by annihilation and 
creation operators,
$b$ and $b^\dagger$ respectively, which obey the usual commutation
relation,
\begin{equation}
\left[b,b^{\dagger}\right] = 1.
\end{equation}
In general the atomic system will be described in terms of a set of matrices.
For example, the two-level system in the RH is described by the SU(2) Pauli 
matrices.

Our method for finding exact isolated solutions for such systems involves
two components.  First, one must choose an appropriate representation for the
atomic matrices and then, crucially, one performs a Bogoliubov transformation
of the operators of the field mode.  The nature of this transformation 
depends upon the type of interaction being considered and, with the 
correct choice of parameters, it leaves the Schr\"{o}dinger
equation in a form that admits exact solution with very simple Ans\"{a}tze.

\subsection{Bogoliubov transformations}
A Bogoliubov transformation is a transformation from one description of a
field mode in terms of the bosonic operators, $b$ and $b^\dagger$, to a 
description in terms of new bosonic operators, $\tilde{b}$ and 
$\tilde{b}^\dagger$, say.  This transformation is canonical so that the
new operators obey the same commutation relation as the old ones, namely
\begin{equation}
\left[\tilde{b},\tilde{b}^\dagger\right] = 1.
\end{equation}
The most general linear Bogoliubov transformation may be viewed as a
rotation plus translation of the original oscillator Hilbert space to
the new oscillator space,
\begin{eqnarray}
\tilde{b} =e^{-i\beta}\rb{1-|\sigma|^2}^{-1/2}\rb{b-\sigma b^\dagger - z} 
		\nonumber \\
\tilde{b}^\dagger =e^{i\beta}\rb{1-|\sigma|^2}^{-1/2}
                   \rb{b^\dagger-\sigma^* b - z^*},
\end{eqnarray}
where $\sigma$ and $z$ are complex numbers describing the amplitudes 
of the rotation and translation respectively. $\beta$ is a simple, and usually
rather unimportant, phase factor. From the outset it is important 
to note the restriction $|\sigma|<1$ in order to preserve the unitarity of the 
transformation.  In the following we consider two specialisations of 
this transformation,
namely a pure translation and a pure rotation.  These transformations may
be very simply related to the familiar coherent and squeezed states of 
quantum optics and it is from this standpoint that we introduce the 
transformations.

\subsection{Coherent bosons}
The usual Glauber coherent states, $\ket{z}$, may be defined as eigenkets of
the single-mode bosonic annihilation operator \cite{ro:gl},
\begin{equation}
b\ket{z} = z \ket{z}, \label{CSdefn}
\end{equation}
where $z$ is a complex number. Such states are readily constructed as 
the following equivalent forms,
\begin{eqnarray}
\ket{z} &=&  e^{-\frac{1}{2}|z|^2} e^{z b^\dagger}\ket{0}
\\ \label{CSexp3}
&=& e^{\rb{zb^\dagger - z^* b}}\ket{0}, \label{CSexp4}
\end{eqnarray}
where we have normalised the coherent state such that $\langle z
\ket{z} = 1$. The exponential operator in Eq. (\ref{CSexp4}) is
denoted as follows,
\begin{equation}
D\left( z \right) \equiv e^{\rb{zb^\dagger - z^* b}}, \label{Dispop}
\end{equation}
and is called the displacement operator.  It is a unitary operator and we 
may readily use it to perform a unitary
transformation of the field operators,
\begin{eqnarray}
D\left( z \right) b D^\dagger\left( z \right) = b - z \equiv a \nonumber \\
D\left( z \right) b^\dagger  D^\dagger\left( z \right) =
b^\dagger - z^* \equiv a^\dagger \label{Dtransf}
\end{eqnarray}
The operators $D\rb{z}$ form a representation of the Weyl (or Heisenberg-Weyl)
group when multiplied by a trivial phase factor $\exp \rb{i \phi}$, with $\phi$
real.
The operators $a$ and $a^\dagger$ obey the same commutator
relation as the original operators, and thus we see this transformation
to be a Bogoliubov transformation of the type described as a pure 
translation above.  Equations (\ref{CSdefn}) and (\ref{Dtransf}) 
clearly imply
\begin{equation}
a\ket{z}=0,
\end{equation}
from which we see that the operator $a$ annihilates the coherent state
$\ket{z}$.  Thus $\ket{z}$ may be considered as the vacuum state
of the $a$-type bosons, and we rewrite it accordingly as 
$\ket{0;z}\equiv \ket{z}$,
\begin{equation}
a \ket{0;z} = 0.
\end{equation}
We shall call these $a$-type bosons ``coherent bosons'' and write their
number states as $\ket{n;z}$, such that 
$a^\dagger a \ket{n;z} = n\ket{n;z}$.

\subsection{Displaced harmonic oscillator}
The simplest application of the coherent bosons is to the displaced harmonic
oscillator,
\begin{equation}
H_\mathrm{D} = \frac{1}{2}\rb{x+\sqrt{2}\lambda}^2 + \frac{1}{2}p^2,
\end{equation}
in which the centre of the oscillator is shifted by an amount 
$-\sqrt{2}\lambda$.
Introducing the harmonic oscillator operators via
\begin{eqnarray}
x \equiv \frac{1}{\sqrt{2}}\left( b^\dagger + b\right), \nonumber
\\ p \equiv \frac{i}{\sqrt{2}}\left(b^\dagger - b\right),\label{xpbbd}
\end{eqnarray}
the  Hamiltonian reads
\begin{equation}
H_\mathrm{D} =  b^{\dagger}b + \lambda
\left(b^{\dagger} + b \right) + \frac{1}{2} + \lambda^2.   \label{shiftHam2}
\end{equation}
By performing a Bogoliubov transformation of the original bosonic
operators to a new set of coherent bosons, $a^\dagger$ and $a$, such that
\begin{equation}
a \equiv b + \lambda, ~~ a \equiv  b^\dagger + \lambda ,
\label{bctran}
\end{equation}
we may rewrite the Hamiltonian of Eq. (\ref{shiftHam2}) in the form
\begin{equation}
H_\mathrm{D} =  a^\dagger a +\frac{1}{2}
\end{equation}
The eigenstates of this Hamiltonian are thus clearly seen to be
the number states of the $a$-type bosons, with corresponding
eigenenergies $E_n = n +\frac{1}{2}$. 

\subsection{Squeezed bosons\label{QSQ}}

Following Bishop and Vourdas \cite{bi:v2} we construct the most general 
squeezed state, $\ket{z; \rho, \theta, \beta}$, by acting 
upon the bosonic vacuum $\ket{0}$ first  with the displacement
operator $D\left(z\right)$ of Eq. (\ref{Dispop}) and then
with the pure squeezing operator $S\left( \rho, \theta, \beta \right)$,
\begin{equation}
\ket{z; \rho, \theta, \beta} = S\left( \rho, \theta, \beta \right)
D \left( z\right) \ket{0},
\end{equation}
The squeezing operator is given by
\begin{equation}
S\left( \rho, \theta, \beta \right) \equiv \exp \left(
-\frac{1}{4}\rho e^{-i\theta} {b^\dagger}^2 + \frac{1}{4}\rho
e^{i\theta} b^2 \right) \exp \left( i \beta b^\dagger b \right),
\end{equation}
where $\rho, \theta, \beta $ are real parameters.  It is a unitary operator, 
$S^\dagger S = 1$, and provides a
representation of the group  SU(1,1).
Using a relationship given by Perelomov \cite{pe:re},
we are able to write the squeezing operator in the equivalent form
\begin{equation}
S \left( \sigma, \beta \right) = \exp \left( \frac{1}{2} \sigma
{b^\dagger}^2\right) \left( 1 - |\sigma|^2\right)^{b^\dagger b /2
+ 1/4} \exp \left( -\frac{1}{2} \sigma^* b^2\right) \exp \left(
i \beta b^\dagger b \right), \label{SQSeqform}
\end{equation}
where $\beta$ is the same real parameter as above, and $\sigma$ 
is a complex number with
modulus $|\sigma| < 1$, given by 
$\sigma \equiv - e^{-i\theta} \tanh \rb{\frac{1}{2}\rho}$.
Using this expression, we can use the
squeezing operator to make unitary transformations of the bosonic
annihilation and creation  operators,
\begin{eqnarray}
S\left( \sigma, \beta \right) b S^\dagger \left( \sigma, \beta
\right) &=& e^{-i\beta} \left( 1 - |\sigma|^2\right)^{-1/2} \left(
b - \sigma b^\dagger\right) \equiv c, \nonumber \\ S\left( \sigma,
\beta \right) b^\dagger S^\dagger \left( \sigma, \beta \right) &=&
e^{i\beta} \left( 1 - |\sigma|^2\right)^{-1/2} \left( b^\dagger -
\sigma^* b\right) \equiv c^\dagger. \label{SQcdefn}
\end{eqnarray}
The operators $c$ and $c^\dagger$ satisfy the commutation relation
$\left[c, c^\dagger \right]=1$ and thus the transformation $b,
b^\dagger \rightarrow c, c^\dagger$ is a Bogoliubov transformation of
the rotation type.
From Eq. (\ref{SQcdefn}), it follows that for any
function of $f \left(b , b^\dagger \right)$
\begin{equation}
S f \left( b, b^\dagger\right) S^\dagger = f\left( c,c^\dagger
\right) \leftrightarrow S f \left( b, b^\dagger \right) = f \left(
c, c^\dagger \right) S.\label{SQftran}
\end{equation}
Equation (\ref{SQftran}) implies that $S b = c S$ and hence
$\ket{z; \sigma \beta}\equiv \ket{z;\rho\theta\beta}$ are 
eigenstates of the annihilation operator $c$,
\begin{eqnarray}
c \ket{z; \sigma \beta} &=& c S\left(\sigma, \beta\right) \ket{z}
\nonumber \\ &=& S\left( \sigma, \beta\right) b
\ket{z} \nonumber
\\ &=& z \ket{z;\sigma\beta}.
\end{eqnarray}
If we consider the squeezed vacuum $S\ket{0} = \ket{0; \sigma,
\beta}= \ket{0; \sigma}$, we see that it is independent of
$\beta$ and that
\begin{equation}
c\ket{0; \sigma} = 0.
\end{equation}
The number states of the $c$-type bosons are denoted
$\ket{n; \sigma \beta}$, such that $c^\dagger c \ket{n; \sigma \beta} =
n\ket{n; \sigma \beta}$.  We call the $c$-type bosons ``squeezed'' bosons.

\subsection{Squeezed harmonic oscillator}
In position representation the squeezed harmonic oscillator has the form
\begin{equation}
H_\mathrm{S} = \frac{1}{2}\rb{1+2\lambda} x^2 
  +\frac{1}{2}\rb{1-2\lambda}p^2,
\end{equation}
where the real parameter $\lambda$ determines the degree of 
squeezing, with the restriction that  $|\lambda|<\frac{1}{2}$.
Translating this into the standard bosonic representation defined by 
Eq. (\ref{xpbbd}) we have
\begin{equation}
H_\mathrm{S} =  b^\dagger b +\frac{1}{2} 
  + \lambda \rb{{b^\dagger}^2 + b^2}.
\label{sqHam2}
\end{equation}
We introduce squeezed $c$-type bosons defined by
\begin{equation}
c^\dagger = \frac{b^\dagger + \sigma b}{\sqrt{1
-\sigma^2}},~~~~~c = \frac{b + \sigma
b^\dagger}{\sqrt{1-\sigma^2}}, \label{squbos1}
\end{equation}
and leave $\sigma$ real but undetermined for the moment.
Making these substitutions into Eq. (\ref{sqHam2}), we have
\begin{equation}
H_\mathrm{S} = \frac{1}{\rb{1-\sigma^2}} \left\{
 \left[-\sigma + \lambda + \lambda\sigma^2 \right]\rb{c^2 +
{c^\dagger}^2} + \rb{\sigma^2+1-4\lambda\sigma}
\rb{c^\dagger c + \frac{1}{2}}\right\}.
\end{equation}
We eliminate the first term in this Hamiltonian by choosing
\begin{equation}
-\sigma + \lambda + \lambda\sigma^2 = 0 \label{sigdet},
\end{equation}
giving, as one of the two solutions,
\begin{equation}
\sigma = \frac{\rb{1-\Omega}}{2\lambda};~~~~\Omega =
\sqrt{1-4\lambda^2}.
\end{equation}
With this choice, the Hamiltonian becomes
\begin{equation}
H_\mathrm{S} =  \left\{c^\dagger
c + \frac{1}{2}\right\}\Omega.
\end{equation}
The eigenstates of this Hamiltonian are clearly the number states
of the squeezed $c$-type bosons, with eigenenergies
\begin{equation}
E_{n} =  \left\{ n +
\frac{1}{2}\right\}\Omega.
\end{equation}
We note that the other solution of Eq. (\ref{sigdet}) with
 $\sigma = \frac{\rb{1+\Omega}}{2\lambda}$ leads to the unphysical
oscillator with $H_\mathrm{S} = -\rb{c^\dagger c + \frac{1}{2}}\Omega$,
and since this Hamiltonian does not have square-integrable solutions,
we discard it.

\section{Application to the Rabi Hamiltonian\label{RHapp}}

The Rabi Hamiltonian (RH) describes a two-level atom interacting with 
a single mode of quantised electromagnetic radiation via a dipole interaction
\cite{al:eb}.  It is usually written in the form
\begin{equation}
H_{\mathrm{Rabi}} = \frac{1}{2} \omega_{0} \sigma_z + \omega
b^{\dagger}b + g\left(b^{\dagger} + b \right)\left(\sigma_{+} +
\sigma_{-}\right),           \label{RH1}
\end{equation}
where $\omega_0$ is the atomic level splitting, $\omega$ is the
frequency of the boson mode and $g$ is the coupling strength of
the atom  to the field.  The two-level atom 
is described by the Pauli pseudo-spin
operators, which satisfy the SU(2) commutation relations
\begin{equation}
\left[\sigma_k, \sigma_l\right]=  2 i \varepsilon_{klm}\sigma_m,
\end{equation}
where $k,l,m \in \left\{ x,y,z\right\}$ with $k \ne l$ and $\varepsilon_{klm}$
is the anti-symmetric Levi-Civita symbol.  We have defined
the raising and lowering operators as
\begin{equation}
\sigma_+ \equiv \sigma_x + i \sigma_y,~~\sigma_- \equiv \sigma_x - i \sigma_y.
\end{equation}
It is convenient to rescale the Hamiltonian as 
$H_\mathrm{Rabi} = \omega \tilde{H}_\mathrm{Rabi}$, where
\begin{equation}
\tilde{H}_{\mathrm{Rabi}} = \tilde{\omega}\sigma_z + b^{\dagger}b + 
\lambda\left(b^{\dagger} + b \right)\sigma_x,\label{RH2}
\end{equation}
and $\tilde{\omega} \equiv \frac{\omega_0}{2 \omega}$ and
$\lambda \equiv \frac{2g}{\omega}$.  
There is a conserved parity $\Pi$ associated with the Hamiltonian,
\begin{eqnarray}
\Pi &\equiv& \exp \left[i \pi \left(b^\dagger b +
\frac{1}{2}\left(\sigma_z + 1\right)\right) \right] \nonumber \\
&=& - \sigma_z \cos \left( \pi b^\dagger b \right), \label{parity}
\end{eqnarray}
such that $\left[H_{\mathrm{Rabi}},\Pi\right]=0$.  The parity operator 
$\Pi$ has two eigenvalues, $\pi=\pm 1$.
The RH is not known to be integrable, but isolated exact solutions 
do exist.  Here we use the technique outlined above to find these Juddian
solutions.

In order to do this we first require an appropriate matrix representation
for the Pauli matrices, which  for this model is one 
in which $\sigma_x$ is diagonal.  We shall use
\begin{equation}
\sigma_x = \left[\begin{array}{lr}1 & 0\\0 & -1 \end{array} \right],~~
\sigma_y = \left[\begin{array}{lr}0 & i\\-i & 0 \end{array} \right],~~
\sigma_z=\left[\begin{array}{lr}0 & 1\\1 & 0 \end{array} \right]. 
\end{equation}

In terms of the two-component 
wavefunction, $\ket{\Psi} = {\ket{\Psi_1}\choose \ket{\Psi_2}}$,
the time-independent Schr\"{o}dinger equation for the system,
$\tilde{H}_\mathrm{Rabi}\ket{\Psi} = E\ket{\Psi}$, then reads
\begin{eqnarray}
\tilde{\omega}\ket{\Psi_2} + \rb{ b^\dagger b + \lambda
\rb{b^\dagger + b} - E}\ket{\Psi_1} &=& 0  \nonumber \\
\tilde{\omega}\ket{\Psi_1} + \rb{b^\dagger b - \lambda
\rb{b^\dagger + b} - E}\ket{\Psi_2} &=& 0. \label{BReqns1}
\end{eqnarray}
We now make the Bogoliubov transformation to the coherent bosons, 
$a^\dagger$ and $a$, specified by
\begin{equation}
a^\dagger = b^\dagger - \lambda,~~a = b - \lambda.
\end{equation}
The vacuum state of these bosons is the coherent state $\ket{\lambda}$.  
It should be noted that this choice of transformation
may be intuited from considering the $\tilde{\omega}=0$ limit of
the Hamiltonian, where the same transformation is used to 
solve the model exactly in this limit, which is essentially equivalent to
the displaced oscillator considered earlier.  With this 
transformation Eqs. (\ref{BReqns1}) become
\begin{eqnarray}
\tilde{\omega}\ket{\Psi_2} + \left\{ a^\dagger a
      + 2\lambda \rb{a^\dagger + a} + 3\lambda^2 - E\right\}\ket{\Psi_1} 
							= 0 \nonumber\\
\tilde{\omega}\ket{\Psi_1} + \left\{ a^\dagger a
      - \lambda^2 - E\right\}\ket{\Psi_2} = 0,
\end{eqnarray}
where the kets $\ket{\Psi_{1,2}}$ are now in the transformed
representation.  For these kets we choose the Ansatz
\begin{eqnarray}
\ket{\Psi_1} &=& \sum_{n=0}^{N-1} p_n \ket{n;\lambda} 
 =  \sum_{n=0}^{N-1} p_n \frac{\rb{a^\dagger}^n}{\sqrt{n!}}\ket{0;\lambda} 
 = P_{N-1}\rb{a^\dagger}\ket{0;\lambda}; \nonumber\\
\ket{\Psi_2} &=& \sum_{n=0}^{N} q_n \ket{n;\lambda}
 = \sum_{n=0}^{N} q_n\frac{\rb{a^\dagger}^n}{\sqrt{n!}}\ket{0;\lambda} 
 = Q_{N}\rb{a^\dagger}\ket{0;\lambda};
\label{BRAnsatz2}
\end{eqnarray}
where $\ket{n;\lambda}$ are number states of the coherent bosons,
$a^\dagger a \ket{n;\lambda} =
n\ket{n;\lambda}$, and we have introduced the polynomials 
$P_{N-1}$ and $Q_{N}$ of order $N-1$ and $N$ respectively.  
Making these substitutions we have
\begin{eqnarray}
\tilde{\omega}\sum_{n=0}^{N} q_n \ket{n;\lambda} &+&
\sum_{n=0}^{N-1}p_n \rb{ n + 3 \lambda^2 - E}
\ket{n;\lambda} \nonumber \\&+& 2 \lambda \sum_{n=0}^{N-1}p_n \sqrt{n+1}
\ket{n+1;\lambda} + 2 \lambda \sum_{n=1}^{N-1}
p_n \sqrt{n} \ket{n-1;\lambda} = 0,
\nonumber\\
\tilde{\omega} \sum_{n=0}^{N-1} p_n \ket{n;\lambda} &+&
\sum_{n=0}^{N} q_n \rb{ n - \lambda^2 - E}
\ket{n;\lambda} = 0.
\end{eqnarray}
Considering the highest number state, $\ket{N;\lambda}$, in the
second of these equations, we see that for this equation to hold
we require
\begin{equation}
\rb{ N - \lambda^2 - E}q_N = 0.
\end{equation}
Since $q_N \ne 0$ by Ansatz, we obtain a determination of the energy
\begin{equation}
E = N - \lambda^2. \label{baselines}
\end{equation}
This equation identifies the Juddian baseline energies, along which the
Juddian solutions lie.  Comparing the coefficients of the
remaining number states gives us $2N+1$ linear equations for the
$2N+1$ coefficients $(p_m, 0 \le m \le N-1)$ and $(q_k, 0 \le k
\le N)$.  To obtain non-trivial solutions, we clearly require the
determinant of this equation set to be zero. This gives the
compatibility condition, providing the locations of the Juddian
points.  The first two conditions ($N=1,2$) have the explicit forms
\begin{eqnarray}
\tilde{\omega}^2 + 4 \lambda^2 =1,~~~~~~~~~~~~~~~~~~~\mathrm{for} ~N=1, \\
\tilde{\omega}^4 +  \rb{12 \lambda^2 - 5} \tilde{\omega}^2 + 32
\lambda^4 - 32 \lambda^2 + 4 =0, ~~~~~~~~~~~\mathrm{for} ~N=2,
\end{eqnarray}
as have been given by K\`{u}s and Lewenstein \cite{ku:le}.
Thus, for a given $N$, we have a polynomial of $N$th order in $\lambda^2$ and
$\tilde{\omega}^2$. Each of these has $N$ roots for $\lambda^2$ in terms of
$\tilde{\omega}^2$, which all turn
out to be real, thus giving the location of $N$ Juddian
solutions.  Before we look at these results, it is of interest to consider 
the other possible type of finite
Ansatz at the Juddian points.  These are found by using the coherent
bosons
\begin{equation}
a^\dagger = b^\dagger + \lambda,~~a = b + \lambda,
\end{equation}
and interchanging the roles of $\ket{\Psi_1}$ and $\ket{\Psi_2}$.

\subsection{Results}

By solving the complementary conditions we have calculated the first
ten Juddian points for the resonant RH. These are
displayed in Table \ref{1pJuddtab}, listed to 10 decimal places.

%Table was here

The location of these Juddian points in the energy schema of the
Hamiltonian is displayed in Figure {\ref{1pJpoints}, where the schema was
obtained by approximate numerical diagonalisation via a
standard configuration-interaction method, using a basis
size of the lowest 101 harmonic oscillator states \cite{ce:th}.
Also plotted are the
Juddian baselines from Eq. (\ref{baselines}).

%figure was here

From Fig. \ref{1pJpoints} we see that the Juddian points occur at the level
crossings in this diagram.  Thus we see that they occur when two
solutions of different parity $\Pi$ become degenerate in energy, and this
degeneracy is the key to the existence of the Juddian
solutions.  The coherent-boson number states $\ket{n;\lambda}$ are
not eigenstates of $\Pi$, and thus the Ansatz
(\ref{BRAnsatz2}) is not of definite parity.  It is precisely
because we can construct wavefunctions of mixed parity that allows
us to find such simple Ans\"{a}tze at the Juddian points.

We are now able to make explicit the connection between 
this method and that used by K\`{u}s and Lewenstein \cite{ku:le} in 
investigating the RH. They worked in the Bargmann representation 
\cite{ba:rg}, in which the bosonic operators are represented by
\begin{equation}
b^\dagger \rightarrow z;~~~~~~ b \rightarrow \frac{d}{dz}.
\end{equation}
and postulated the following forms for the two components of the
wavefunction:
\begin{eqnarray}
\Psi_1\rb{z} = e^{-|z|^2/2}\langle z | \Psi_1\rangle 
= \widetilde{P}_{N-1}\rb{z} e^{\lambda z}, \nonumber \\
\Psi_2\rb{z} =e^{-|z|^2/2}\langle z | \Psi_2\rangle 
= \widetilde{Q}_{N}\rb{z} e^{\lambda z},\label{KL}
\end{eqnarray}
where $\widetilde{P}_{N-1}\rb{z}$ and $\widetilde{Q}_{N}\rb{z}$ are 
polynomials in $z$ of order $N-1$ and
$N$ respectively. Bearing in mind the form of the coherent state
(\ref{CSexp3}), these 
wavefunctions are simply seen to be of the form of polynomials in the 
bosonic creation operator, $b^\dagger$, acting upon a coherent state of 
amplitude $\lambda$.  In our
Ansatz (\ref{BRAnsatz2}), we have the same coherent 
state but now being acted upon 
by polynomials in 
$a^\dagger = \rb{b^\dagger - \lambda}$, which shares a closer connection to
the coherent state than $b^\dagger$.

The polynomials of K\`{u}s and Lewenstein, are simply related to those 
of Ansatz (\ref{BRAnsatz2}) by  $\widetilde{P}_{N}\rb{z} = P_N\rb{z-\lambda}$.
In the present case where
we have only used displacements of the boson mode, the difference 
between the two approaches
is thus minimal.  However, this is not the case when we require the use of 
squeezed bosons.  Generally, an Ansatz posited in the 
squeezed representation would contain polynomials of 
the form $P_N\rb{c^\dagger}$, where $c^\dagger$ is the 
creation operator of the squeezed bosons.  The analogous Ansatz 
to Eq. (\ref{KL}) would still contain a polynomial in $z$,
 $\widetilde{P}_N\rb{z}$ say.  
If we assume the simplest type of squeezing and write 
$c^\dagger = \rb{b^\dagger + \sigma b}/\rb{\sqrt{1-\sigma^2}}$ as in Eq.
(\ref{squbos1}), then the K\`us and Lewenstein polynomial can be written
\begin{eqnarray}
\widetilde{P}_N\rb{z} = P_N\rb{ \frac{z + 
  \sigma \frac{d}{dz}}{\sqrt{1-\sigma^2}}},
\end{eqnarray}
which, crucially, contains both $z$ and its derivative, and although formal 
relationship do exist between the polynomials 
of the two methods, these relationships are generally not trivial,
especially if one considers the more general form of the Bogoliubov 
transformation.  So the Ans\"{a}tze of the two methods are seen to be
significantly different, and we conjecture that the one described 
here has several advantages which we shall discuss in the conclusion.

\section{Conclusions}

We have presented a method for finding isolated exact solutions of a class of
non-adiabatic models, of the type frequently used in quantum optics
and related fields.

Compared with the original approaches of Judd and Reik, 
the above method is more transparent and considerably simpler, 
advantages that it 
shares with the technique of K\`{u}s and Lewenstein.  However, we 
believe that the use of 
transformed bosons is more obviously physically meaningful than the 
use of wavefunctions in Bargmann space, especially given the connection 
of these bosons to the coherent and squeezed states, 
so important in quantum optics.

As an example of the use of this
technique, we have applied it to the Rabi Hamiltonian and obtained in a 
simple fashion the known Juddian solutions of this model.
In this example, we have used the coherent bosons to obtain
Juddian solutions for a problem with an interaction of the type 
$\lambda\rb{b^\dagger+ b}\sigma_x$.  It is hopefully now clear 
how one may apply
this method to further problems containing the same type of interaction.  We
have not as yet mentioned the application of the squeezed bosons in performing
this kind of calculation. This second type of Bogoliubov transformation
is useful in finding
Juddian solutions of models containing two-photon type interactions.
An obvious example 
is the two-photon Rabi Hamiltonian \cite{ng:lo}, which has the Hamiltonian
\begin{equation}
H = \tilde{\omega}\sigma_z + b^{\dagger}b + 
\lambda\left({b^{\dagger}}^2 + b^2 \right)\sigma_x.\label{TPRH}
\end{equation}
Using squeezed bosons we are able to obtain a set of Juddian solutions for
this model and these results will be discussed in a future 
publication.

Due to the intuitive nature and simplicity of this technique it
is easy to extend to other systems.  For example, in view of their mode
of construction we expect that our displaced and squeezed coherent states will
be of particular use in any quantum field theory that has underlying 
dynamical symmetry of the Weyl group or the SU(1,1) group, or to which
the (inhomogeneous or homogeneous) Bogoliubov transformation may be profitably 
applied.  The obvious group-theoretical foundations of the technique also 
point the way to other approximations, since, for example in the squeezed 
(two-photon) case, SU(1,1) is not the only relevant group.  Thus, the 
three-dimensional Lorentz group SO(2,1), which is the group of rotations in 
three-dimensional Minkowski space with two space and one time dimensions, 
is locally isomorphic to SU(1,1).  Similarly, both the groups 
SL(2,$\mathbb{R}$) of real second order matrices with unit determinant and 
the symplectic group Sp(2,$\mathbb{R}$) are also locally isomorphic to SU(1,1).

One may also readily generalise the current approach
for the two-level models involving linear or quadratic interactions with
a single boson (or canonical quantum mode) to the corresponding case of 
linear or bilinear interactions involving several distinct bosons or modes.
For the linear models involving only displacements this is essentially trivial.
However, for models involving squeezing, in the case of $n$ bosons or modes
the various bilinear products of operators $b_i^\dagger b_j^\dagger$, 
$b_i b_j$ and $b_i^\dagger b_j$, $i,j =1,2,\ldots, n$ now form a realisation of
the higher symplectic algebra Sp(2n, $\mathbb{R}$).  As before one can simply
construct a unitary representation of this group by exponentiating the
skew-adjoint operators in the algebra.  For example, Bishop and Vourdas 
\cite{bi:v3} have shown explicitly how to construct the most general two-mode
squeezed states associated with a unitary representation of the group 
Sp(4,$\mathbb{R}$).  Once again such states are the ordinary coherent states
with respect to the new destruction operators $c_1$ and $c_2$, which are
themselves general linear Bogoliubov transformations of the original
destruction operators $b_1$, $ b_2$ and their Hermitian-conjugate creation
operators $b_1^\dagger$, $b^\dagger_2$.  The Sp(4,$\mathbb{R}$) algebra has 
various subalgebras corresponding to different sorts of linear pairing terms.
For example, whereas the single-mode paring operators 
$K_+^{(i)} \equiv \frac{1}{2} \rb{b_i^\dagger}^2$;
$K_-^{(i)} \equiv \frac{1}{2} b_i^2$;
$K_0^{(i)} \equiv \frac{1}{2} b_i^\dagger b_i + \frac{1}{4}$ for $i=1,2$ 
correspond to the so-called $\rb{\frac{1}{4}, \frac{3}{4}}$ representations
 of SU(1,1), the mixed pairing operators 
$L_+ \equiv b_1^\dagger b_2^\dagger$;
$L_- \equiv b_1 b_2$;
$L_0 \equiv \frac{1}{2} \rb{b_1^\dagger b_1 + b_2^\dagger b_2 +1}$ correspond
 to the discrete-series representation of SU(1,1).  By contrast, the mixed
 pairing
operators $J_+ \equiv b_1^\dagger b_2$; $J_- \equiv b_1 b_2^\dagger$; $J_0 
\equiv \frac{1}{2} \rb{b_1^\dagger b_1 - b_2^\dagger b_2}$ correspond to the 
(Schwinger representation of) the angular momentum subalgebra SU(2).

Bishop and Vourdas have shown in a separate publication \cite{bi:v4} how 
squeezed (pair) coherent states can also be used in connection with a 
rather broad class of quantum Lagrangians which include the damped harmonic
 oscillator, and hence with problems involving ``quantum friction'' or
 fluctuation-dissipation phenomena in general.  Within quantum optics for 
example, the quantum theory of lasers and photon detection provide obvious 
applications.  Such problems can now also usefully be extended by our present
treatment to the case of such damped systems coupled to two level atoms.

The possibility of using these solutions as the basis of a 
perturbative approach extends the method away from just the isolated exact 
points to the remainder of the spectrum of the system. The properties of
such an approach are yet to be investigated.  Finally we note that the 
extension to similar single-mode or multi-mode 
systems as considered above coupled to $n$-level atoms with $n \ge 2$ is 
also straightforward in principle.

\section{Acknowledgments}

C. E. acknowledges the financial support of a research studentship
from the Engineering and Physical Sciences Research Council (E.P.S.R.C.) of
Great Britain.

\newpage
\begin{table}[p]
\begin{center}
\begin{tabular}{|c|c|c|}
\hline
$g$ & $E$ & $N$ \\
\hline \hline
0.2165063510 & 0.8125000000 & 1 \\
0.1661640732 & 1.8895580031 & 2 \\
0.4460403578 & 1.2041919969 & 2 \\
0.1400889590 & 2.9215003343 & 3 \\
0.3664714887 & 2.4627945920 & 3 \\
0.6163829153 & 1.4802884071 & 3 \\
0.1234229399 & 3.9390671161 & 4 \\
0.3199075781 & 3.5906365658 & 4 \\
0.5243395120 & 2.9002723045 & 4 \\
0.7582492415 & 1.7002323511 & 4 \\
\hline
\end{tabular}
\caption{The couplings, energies, and $N$, of the
first ten Juddian points of the resonant Rabi Hamiltonian
($\omega = \omega_0 = 1$).\label{1pJuddtab}}
\end{center}
\end{table}
\newpage
\begin{figure}[p]
\vspace{1cm}
\centerline{\includegraphics[height=4in]{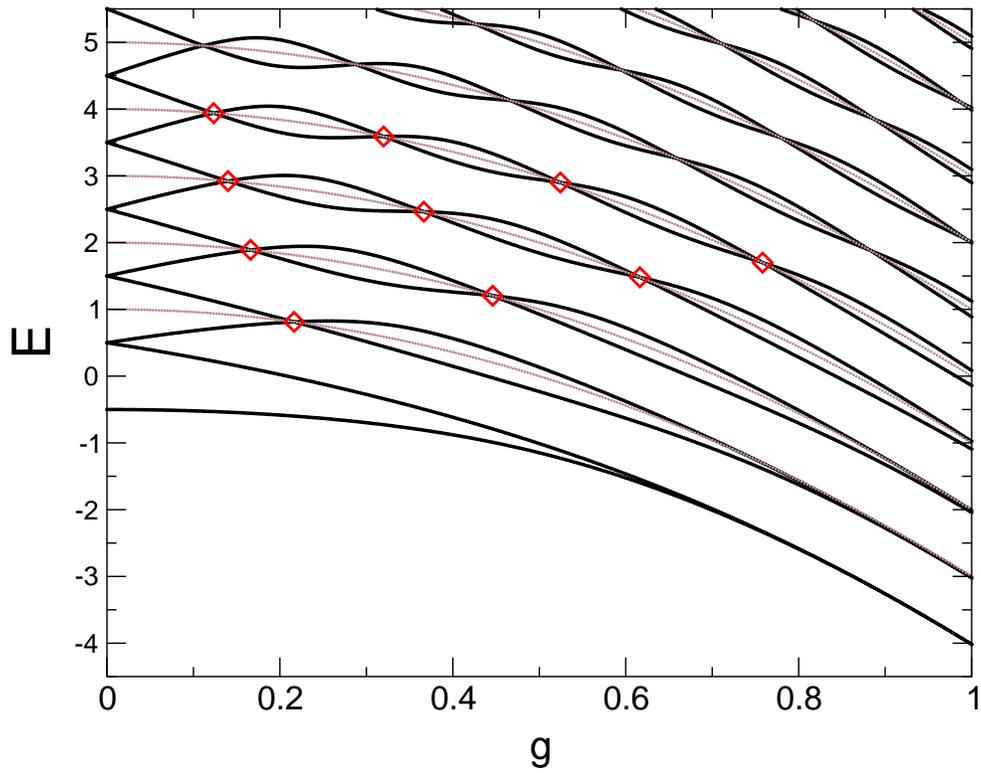}}
\caption{\label{1pJpoints} The first ten Juddian points of the
Rabi Hamiltonian (diamonds).  Also plotted are the energy
levels obtained by numerical diagonalisation (dark lines), 
and the Juddian base-lines (light lines).  The
Hamiltonian is resonant; $\omega = \omega_0 = 1$.  }
\end{figure}
\end{document}